\documentclass[fleqn,usenatbib]{mnras}
\usepackage[T1]{fontenc}
\DeclareRobustCommand{\VAN}[3]{#2}
\let\VANthebibliography\thebibliography
\def\thebibliography{\DeclareRobustCommand{\VAN}[3]{##3}\VANthebibliography}
\usepackage{amstext}
\usepackage{url}
\usepackage[]{times,refname,amsmath,bm}
\bibpunct{(}{)}{;}{a}{}{,}
\usepackage{tabularx}
\usepackage{graphicx,epsfig,color,latexsym}

\usepackage{amssymb, dcolumn, epsf, graphicx, latexsym, mathbbol, slashed, refname, natbib}

\newcommand{\bea}{\begin{eqnarray}}
\newcommand{\eea}{\end{eqnarray}}
\newcommand{\be}{\begin{equation}}
\newcommand{\ee}{\end{equation}}
\newcommand{\rund}[1]{\left(#1\right)}
\newcommand{\vc}[1]{\mbox{\boldmath $#1$}}

\newcommand{\eck}[1]{\left[ #1 \right]}

\defcitealias{er&mao2014}{EM14}

\title[lensed FRB]{The effects of plasma on the magnification and time delay of strongly lensed fast radio bursts}

\author[Er\& Mao]{Xinzhong Er$^{1}$,\thanks{Email: phioen@163.com}, Shude Mao$^{2,3}$ \\	
$^1${South-Western Institute for Astronomy Research, Yunnan University, Kunming, P.R.China}\\
$^2${Department of Astronomy, Tsinghua University, 100084 Beijing, P.R.China}\\
$^3${National Astronomical Observatories, Chinese Academy of Sciences, 20A Datun Road, Chaoyang District, Beijing 100101, P.R.China}\\
}%
\date{Accepted XXX. Received YYY; in original form ZZZ}

\pubyear{2022}

\begin{document}
\label{firstpage}
\pagerange{\pageref{firstpage}--\pageref{lastpage}}
\maketitle

\begin{abstract}
The number of identified Fast Radio Bursts (FRBs) is increasing rapidly with current and future facilities. Strongly lensed FRBs are expected to be found as well, which can provide precise time delays and thus have rich applications in cosmology and fundamental physics. However, the radio signal of lensed FRBs will be deflected by plasma in lens galaxies in addition to gravity. Such deflections by both gravity and plasma will cause frequency dependent time delays, which are different from the dispersion delay and the geometric delay caused by gravitational lensing. Depending on the lensing and plasma models, the frequency-time delay relation of the lensed images can show distinguishing behaviours either between the multiple images, or from the dispersion relation. Such phenomena cannot be neglected in future studies, especially at low radio frequency, as plasma exists in lens galaxies in general. More importantly, such information provides not only a potential way to search for lensed FRBs, but also constraints on the mass and plasma distributions in lens galaxy. In particular, plasma may make the missing central images observable at low radio frequency.
\end{abstract}


\section{Introduction}
Fast Radio Bursts (FRBs) are a new class of extra-galactic millisecond duration radio transients \citep[e.g.][]{lorimer2007,2019A&ARv..27....4P}. FRB signals exhibit a frequency dispersion, usually termed as the `dispersion measure' (DM), which can be used to estimate the free electrons along the line of sight to the FRBs. The large DM observed in some FRBs shows their cosmological origin \citep[e.g.][]{tendulkar2017}, thus the contribution of the DM can come from the Milky Way, the inter-galactic medium and the host galaxy \citep[e.g.][]{yang2017}. The number of identified FRBs has increased dramatically in recent years \citep[e.g.][]{chimecat2021}. With current estimates of the events rate of FRB observation and strong lensing probability, the predicted number of the strongly lensed FRBs by galaxies is also significant, $\sim 10$ per year \citep[e.g.][]{2019PhRvD..99l3517L}. Thanks to high precision time delays, it has been proposed to use the lensed FRBs to study cosmology and fundamental physics \citep[e.g.][]{2017ApJ...847...19D,2019PhRvD..99l3517L,2021A&A...645A..44W}. In order to efficiently search for the lensed FRBs, we need to study the properties of lensed FRBs carefully. It has been noticed that besides gravity, plasma can also cause deflection to the radio signal \citep[e.g.][]{1979Sci...205.1133E,2015PhRvD..92j4031P,2016MNRAS.463L..51H,2018MNRAS.478.4816S,2019PhRvD..99l4001C}. The lensed images will be slightly different in image positions and arrival times depending on the observing bands \citep{er&mao2014} (EM14 in the rest of this paper). The small change due to plasma is usually below $\sim 1$ milli-arcsec in image positions and below second in the arrival time. It is difficult to detect these with the current facilities in cases of lensed QSOs or galaxies, since the intrinsic size and variability timescale of the lensed objects are larger than the changes caused by plasma. However, for lensed FRBs, their short, milli-second durations allow time delays caused by lensing to be easily measured at different frequencies. The frequency-delay relation will be changed by lensing, such a change will not only bias the estimates of the electron density, but also provide a potential method to search for the lensed FRBs. 
In this paper, we adopt the standard $\Lambda$CDM cosmology with parameters based on the results from the Planck collaboration \citep{2020A&A...641A...6P}: $\Omega_\Lambda=0.6847$, $\Omega_m=0.315$, and Hubble constant $H_0=100h$\,km\,s$^{-1}$\,Mpc$^{-1}$ with $h=0.674$. 

\section{Basics}
The basics of gravitational lensing can be found in e.g. \citet{2006glsw.conf....1S}. The thin-lens approximation is adopted, implying that the lensing mass and plasma distribution can be projected on to the lens plane perpendicular to the line-of-sight. We denote the angular diameter distances between the source and the lens as $D_{ds}$, between the source and the observer as $D_s$ and between the lens and the observer as $D_d$. We introduce the angular coordinates $\vc\theta=(\theta_x, \theta_y)$ in the lens plane, and those in the source plane as $\vc\beta=(\beta_x, \beta_y)$. The lens equation with plasma can be written as 
\be
\vc \beta = \vc \theta -\alpha (\vc \theta) 
=\vc \theta - \nabla_\theta\psi_g(\vc\theta) - \nabla_\theta\psi_p(\vc\theta),
\ee
where $\alpha$ is the total deflection angle, $\nabla_\theta$ is the gradient on the image plane, $\psi_g$ is the effective gravitational lens potential, and $\psi_p$ is the effective plasma lensing potential, which is determined by the projected electron density $N_e$
\be
\psi_p(\theta) \equiv \frac{D_{ds}}{(1+z_d) D_d D_s} \frac{\lambda^2}{2\pi} r_e\, N_{\rm e}(\theta) \propto \nu^{-2},
\ee
where $r_e$ is the classical electron radius, and $\lambda$ is the observed wavelength of the signal, and $\nu$ is the frequency. The brightness of the source will be changed by lensing, which is determined by the Jacobian of the lens equation $A$, $\mu=1/{\rm det}(A)$, and $A=\partial{\vc \beta}/\partial{\vc \theta}$.

The usual frequency-time delay relation is given by
\be
t_{\rm DM}\approx 4.16\, {\rm ms}\,
\dfrac{DM}{\nu^2_{}(1+z_d)}.
\label{eq:dm-relation}
\ee
$DM$ is often used to estimate the projected electron density and given in units of pc\,cm$^{-3}$. It has been noticed that such an estimate contains some small biases. For example, the contribution from ions or magnetic field has not been included \citep[][]{2020arXiv200702886K}. Moreover, the plasma lensing effect can also bias the estimate of dispersion measure \citep[][]{er+2020}. For a lensed FRB, DM only accounts for one part of the total time delay, which is composed of three terms,
\be
t_{\nu}(\theta)=\frac{D_t}{c}\eck{\frac{|\theta-\beta|^2}{2} - \psi_g(\theta) +\frac{1}{(1+z_d)}\psi_p(\theta)},
\label{eq:totaldelay}
\ee
where we define $D_t\equiv {D_d D_s(1+z_d)}/{D_{ds}}$.
At a first look, only the third term in Eq.\,\ref{eq:totaldelay} is frequency dependent, i.e. through DM. In reality, the plasma in the lens galaxy will deflect the signal (\citetalias{er&mao2014}). Thus the image positions $\theta$ are frequency dependent as well, which will introduce frequency dependence into the other two delays. In order to see that, we adopt approximations about the deflection angles: $\alpha_{p1,2}$ and $\nabla \alpha_g$ are small, and $\alpha_{g2}- \alpha_{g1}\approx (\alpha_{p2}-\alpha_{p1})\nabla \alpha_{g1}$, where $\alpha_{p1}$ and $\alpha_{p2}$ are the deflection angles caused by plasma at two frequency $\nu_1$ and $\nu_2$ respectively, and $\alpha_{g1}$ is the gravitational deflection angle at frequency $\nu_1$.
Then the time delay in Eq.\,\ref{eq:totaldelay} between $\nu_1$ and $\nu_2$ is approximately 
\begin{align}
\Delta t_{\nu} \approx \dfrac{D_t}{c} &\eck{ \frac{\alpha_{p1}^2 - \alpha_{p2}^2}{2} 
- \alpha_{p2} (\alpha_{p1}-\alpha_{p2}) \nabla \alpha_{g1} }
\nonumber \\
&+ \frac{D_t}{c(1+z_d)}\rund{\psi_p(\nu_1)-\psi_p(\nu_2)},
\label{eq:2delays-nu}
\end{align}
In Eq.\,\ref{eq:2delays-nu}, one can see that besides the dispersion delay, there are extra contributions to the frequency-time delay relation. It has been noticed that the gradient of DM can change the dispersion relation \citep{2016ApJ...821...66L}. The variations in the time delay will introduce errors in pulsar timing array for gravitational wave detection \citep[e.g.][]{2013CQGra..30v4006S,2013CQGra..30v4015S}. Here we present the three contributions to the frequency-time delay difference in turn. 
1) the last term is the dispersion delay, where we assume that the electron density (at the image position) does not change over the frequency. 2) the first term is proportional to $\alpha_{p1}^2-\alpha_{p2}^2$, which can be considered as the geometric time delay caused by the plasma deflection \citep[e.g.][]{FRBplasma1,er+2020}. Unlike the dispersion delay, the geometric delay depends on the frequency to the power of -4, i.e.
\be
\dfrac{c^3 r_e^2}{8 \pi^2D_t}(1+z_d)^2(\nabla N_e)^2 \rund{\frac{1}{\nu_1^4}-\frac{1}{\nu_2^4}}.
\label{eq:2delays-1st}
\ee
The deflection by the plasma lens is opposite to that by gravity \citep{1979Sci...205.1133E,2010MNRAS.404.1790B}, and is stronger at low frequency. 
3) The second term is a mixture term between the plasma and gravitational deflection and it depends on $\nabla\alpha_{g}$, which can be approximated by the lensing shear $\gamma$. Thus the delay caused by the second term can be roughly estimated as $\alpha_{p}^2\gamma\, D_t/c$. The total delay can be complicated. Depending on both the gravitational and the plasma lensing, one of the terms can dominate. For example, at the position of strong shear, the mixture term can be significant. 
It is possible the approximations break down when the gravitational deflection has rapid spatial variations, e.g. close to a sub-halo or stars.
Moreover, the dispersion delay has a weak dependence on the redshift, while the geometric time delay does not. When the lens is located at high-z, one expects that the geometric and mixture term will contribute more to the total time delay.

\section{A mock lensed FRB}
\begin{figure}
\includegraphics[width=10cm]{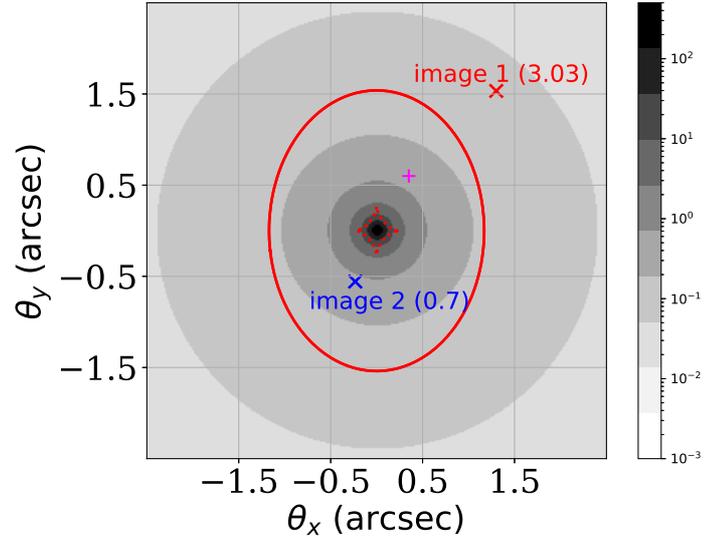}
\caption{
Mock lensed images. Plasma model 2 with $n_0=0.1$\,cm$^{-3}$ is used here. The crosses (plus) mark the positions of the images (source). The numbers in the parentheses are the magnifications. The red solid (dotted) curve is the critical curve (caustics) generated by the lens. 
The grey colour scale presents the time delay difference $t(\nu=1 \, {\rm GHz})-t(\nu=1.5 \, {\rm GHz})$ in units of seconds by Eq.\,\ref{eq:totaldelay}.}
\label{fig:td-map}
\end{figure}

We perform ray-tracing simulations to study the frequency-time delay relation of lensed FRBs. The current observation of FRBs is insufficient to provide a complete redshift distribution. Predictions of the FRB redshift distribution have been given based on the DM-redshift assumption \citep[e.g.][]{2019A&A...621A..91W,chimecat2021,2022MNRAS.509.4775J}. A redshift peak around $\sim0.5$ has been suggested, which coincides with the redshift of the background galaxy in an FRB reminiscent of lensing \citep{2021arXiv211111764X}. We thus adopt this redshift $z_s=0.5534$ as our example. We use a well studied strong lensing system RXJ1131-1231 as our lens \citep{2013ApJ...766...70S} and adopt a Singular Isothermal Ellipsoid (SIE) model \citep{kormann1994} for the total mass distribution (halo and plasma), and place the FRB where two images are generated by the lens with or without plasma (Fig.\,\ref{fig:td-map}). The velocity dispersion of the halo is $\sigma_{v}=330$ km/s with axis ratio $q=0.76$. The redshift of the lens is $z_d=0.295$. Such a lens will generate an Einstein radius of $\theta_E=1''.34$. The time-delay difference between the two images without plasma is about $147$ days, which is much larger than the delay changes between frequencies.


We follow the models in \citetalias{er&mao2014}, in which the details of the plasma lensing can be found
\citep[e.g.][]{2010MNRAS.404.1790B,2020MNRAS.491.5636T}. The electron density models are from extra-galactic observations \citep[e.g.][]{2010ApJ...710L..44G,2013RMxAC..42...14B}, which are within the possible range of the electron density model of the Milky Way \citep{NE2001a,2017ApJ...835...29Y}. 

Two plasma density models are adopted in this study. 1) Model 1 is an exponential profile measured from spiral galaxies, $n(r)=n_0{\rm e}^{-r/r_0}$ with $n_0=2$\,cm$^{-3}$ and $r_0=2$ kpc. 2) Model 2 is a power-law profile measured from elliptical galaxies, $n(r)=n_0(r/r_0)^{-1.25}$ with $n_0=0.1$\,cm$^{-3}$ and $r_0=10$ kpc. The central density of electrons is chosen such that the total mass of the plasma, i.e. free electrons and ionised hydrogen, is significantly smaller than the lens halo mass within the Einstein radius, $\sim 2\%$ for model 1 and $\sim 0.1\%$ for model 2. Even at large radii ($\sim 10$ kpc), the plasma mass of two models is still small. Thus the deflection caused by the gravity of plasma will be neglected currently and left for a future work.

\begin{figure}
\centering
\includegraphics[width=8cm]{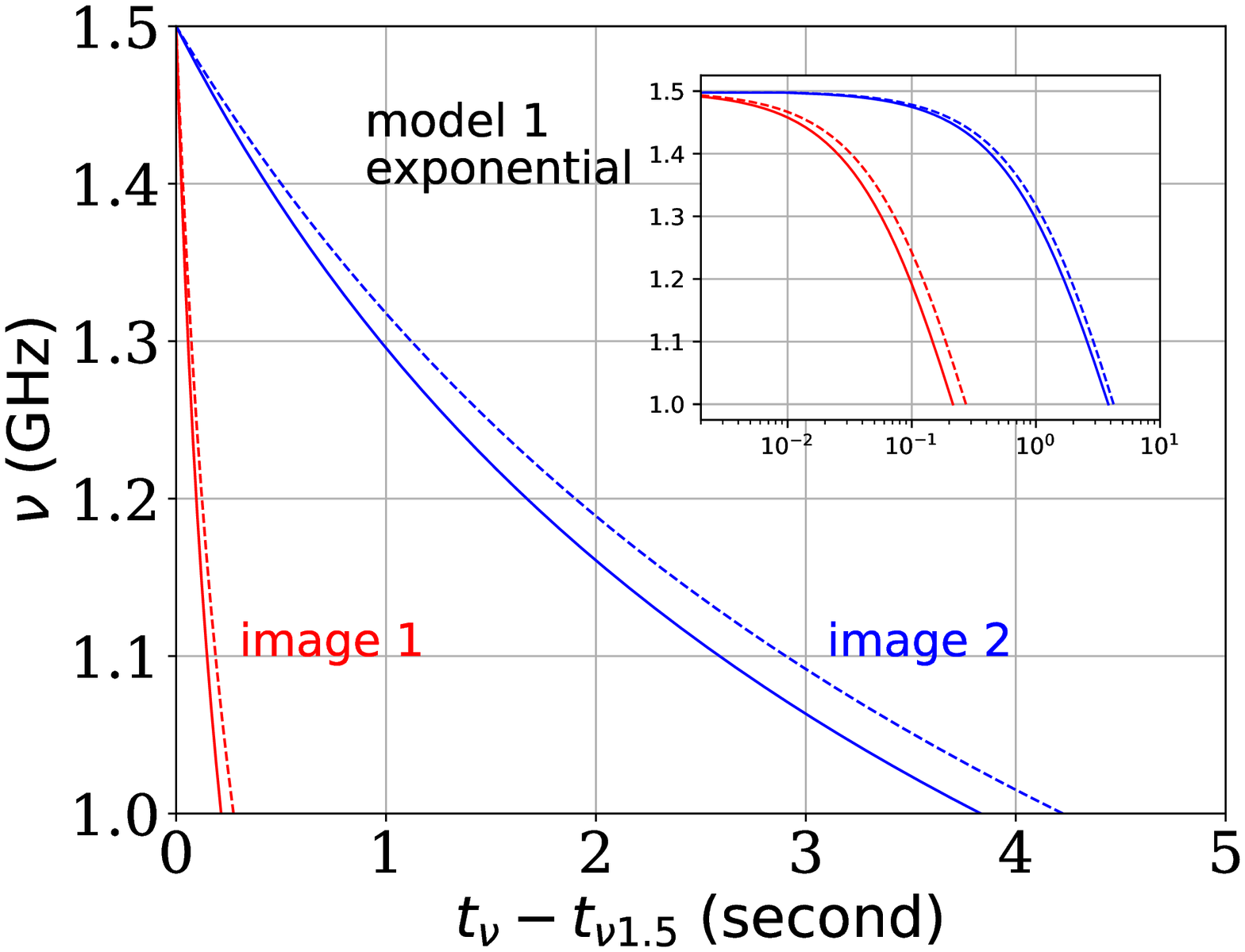}\\
\includegraphics[width=8cm]{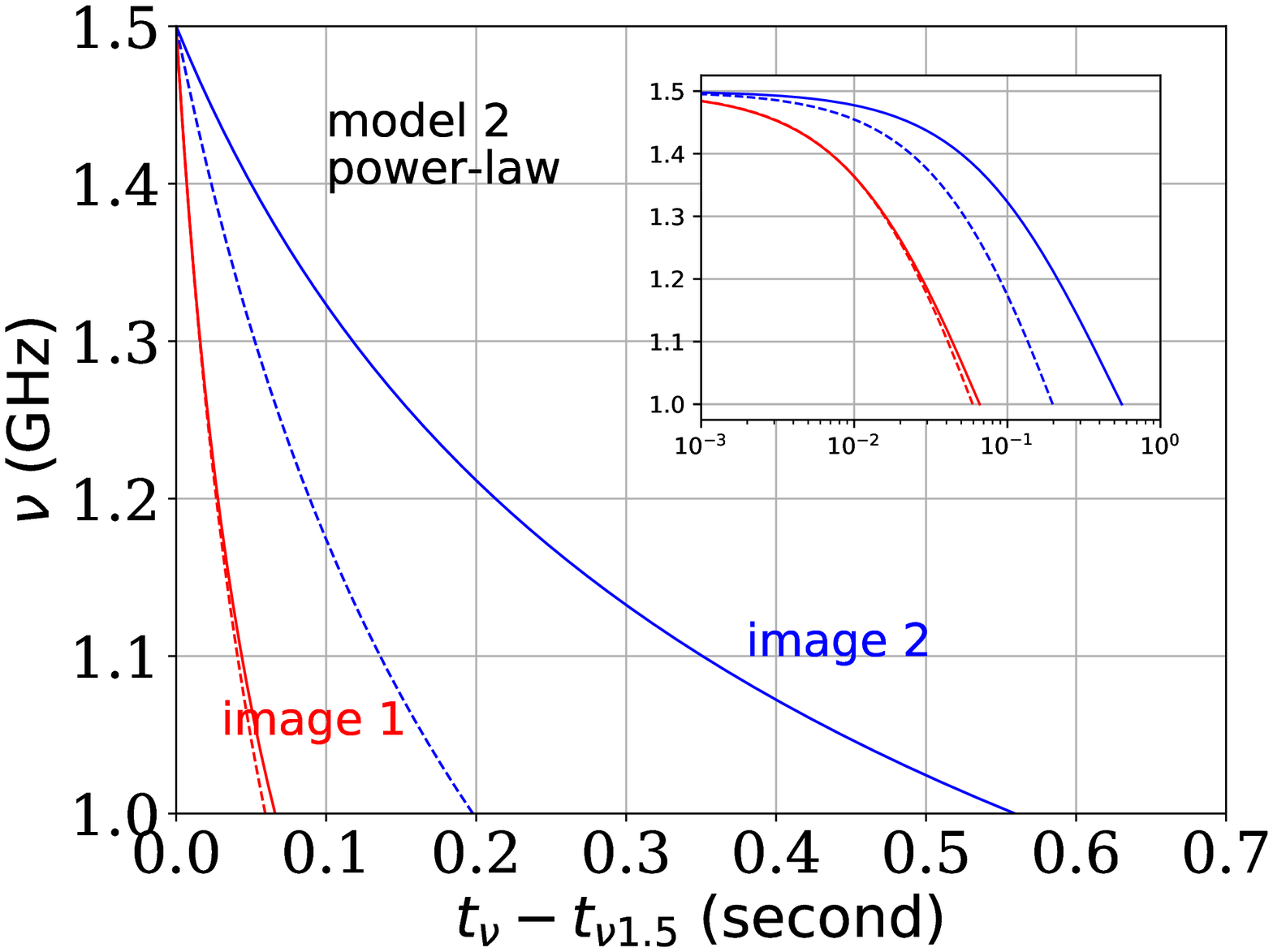}
\caption{The frequency-delay relation of lensed (solid) and unlensed FRBs (dashed). The red (blue) curves show the relation for image 1 (image 2). The delay at high frequency $t_{\nu1.5}$ is subtracted for better visibility. In the top (bottom) panel, we use the plasma model 1 (model 2). For the unlensed ones, the curves represent the dispersion relation of electron density adopted at the position of FRB image in each plasma model. The small panels are the same curves but in log-scale to show more clearly the difference.}
\label{fig:td-frb}
\end{figure}

The image positions changed by the plasma are tiny: for model 1, the changes are $<$ 0.1 milli-arcsec for the bright image and $\sim$0.5 milli-arcsec for the faint image. They are similar for model 2. The time delay on the other hand will show a detectable difference. We calculate the total time delay as a function of frequency (Eq.\,\ref{eq:totaldelay}) in these two plasma models, and compare them with the dispersion delay caused by the uniform plasma density at the corresponding image positions (Fig.\,\ref{fig:td-frb}). In model 1, gravity and plasma slightly decrease the delay at low frequency, since the electron density almost does not change and the density gradient is small. In model 2, the two images show a large difference: the second image (with lower magnification) shows a large delay at low frequency, which roughly agrees with order of magnitude estimate ($\alpha_{p}^2\gamma\,D_t/c$, where $\gamma$ is lensing shear).
This is because the density gradient in model 2 is large, so is the time delay. We show the total time delay difference between 1 and 1.5 GHz using Eq.\,\ref{eq:totaldelay} in Fig.\,\ref{fig:td-map}. Basically when the image is close to the centre of the lens, the time delay is dominated by the lensing effects, and can be large to tens or hundreds of seconds. 

\begin{figure}
\includegraphics[width=10cm]{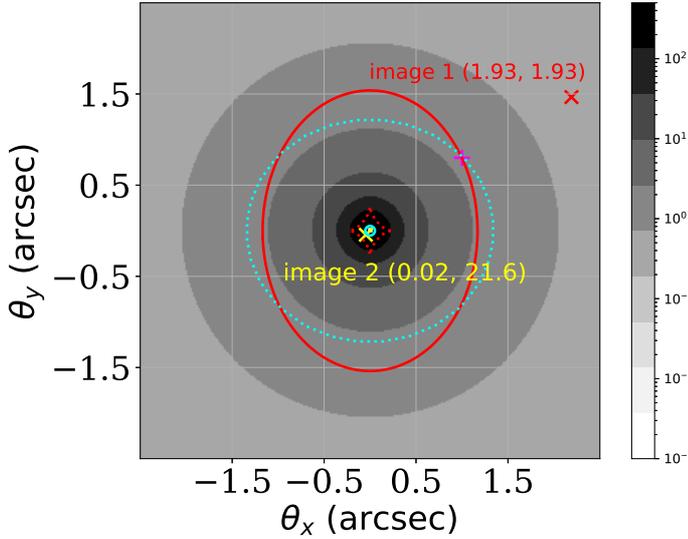}
\caption{
Same as Fig.\,\ref{fig:td-map} but with higher electron density ($n_0=0.5$ cm$^{-3}$) and different image positions. The cyan curve is the extra critical curve (caustics) at $\nu=1$ GHz. The two numbers in the bracket are the magnification generated by gravitational lensing alone and that with additional plasma lensing.
}
\label{fig:td-map2cc}
\end{figure}

The change in magnification depends on the image position as well, i.e. it becomes significant when the image locating near the centre of the lens. In our first example, the changing of magnification by plasma are small ($\sim 1\%$ or even smaller) in both two plasma models. An interesting point is that in gravitational lensing (without plasma), the images near the lens centre will be strongly de-magnified. However, the plasma deflection is opposite to that by gravity. The images near the lens centre by gravitational and plasma lensing will be less de-magnified or even strongly magnified. Such an effect will cause significant frequency variations, for an image at low frequency, the lensed host galaxy may not be observable (de-magnified) but only the lensed FRB. 
We show such an interesting case in an additional example (Fig.\,\ref{fig:td-map2cc}). The plasma model 2 with a higher electron density ($n_0=0.5$\,cm$^{-3}$) is used. A second critical curve (cyan) is generated by the plasma. The magnification of the central image can be dramatically changed by plasma from un-observable to bright \citep[e.g.][]{FRBplasma1}. In this example, the magnification has been changed from $0.02$ to $21.6$. If seen, this will be a dramatic confirmation of plasma lensing.
%

\begin{figure}
\includegraphics[width=7cm]{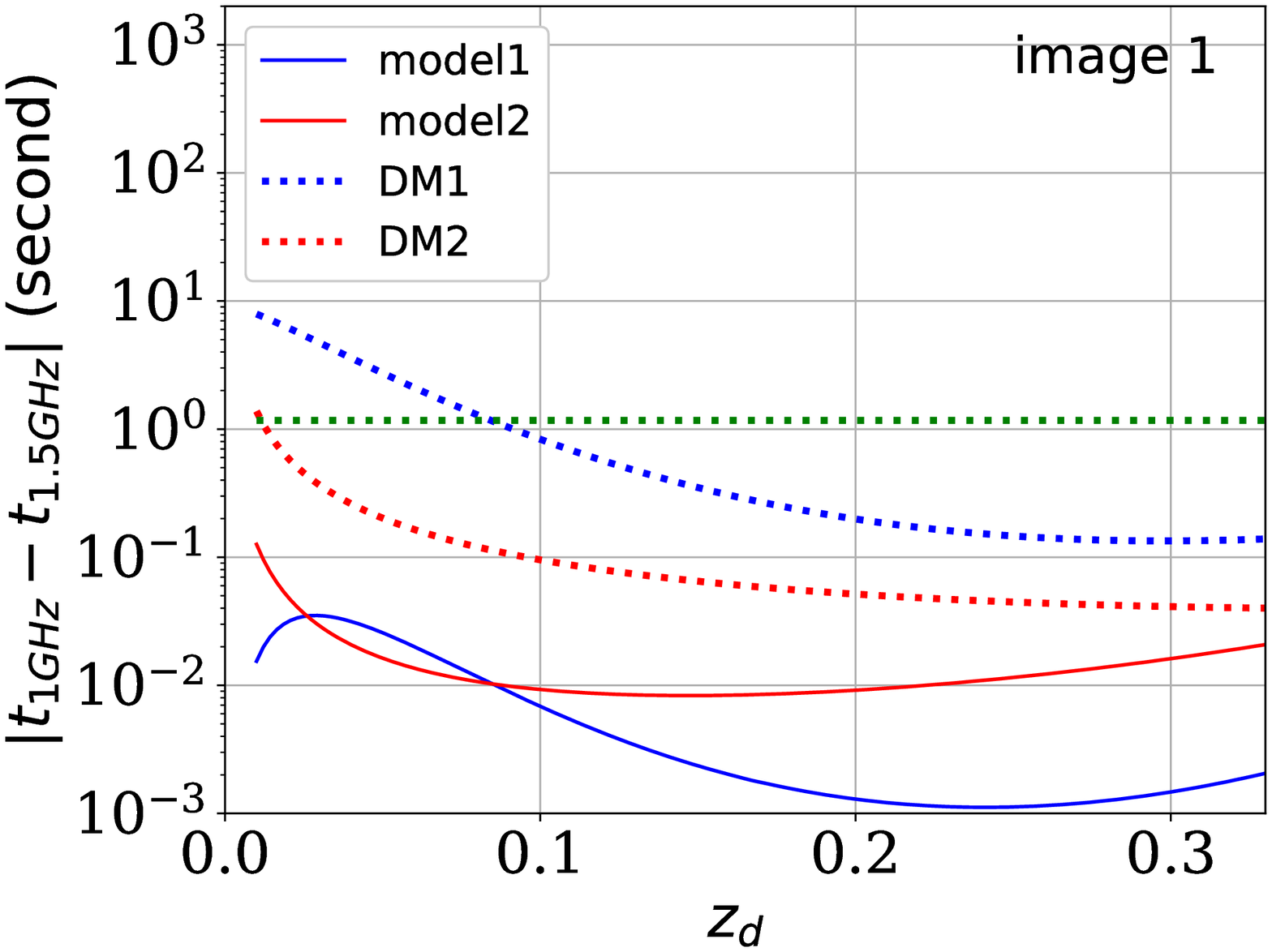}
\includegraphics[width=7cm]{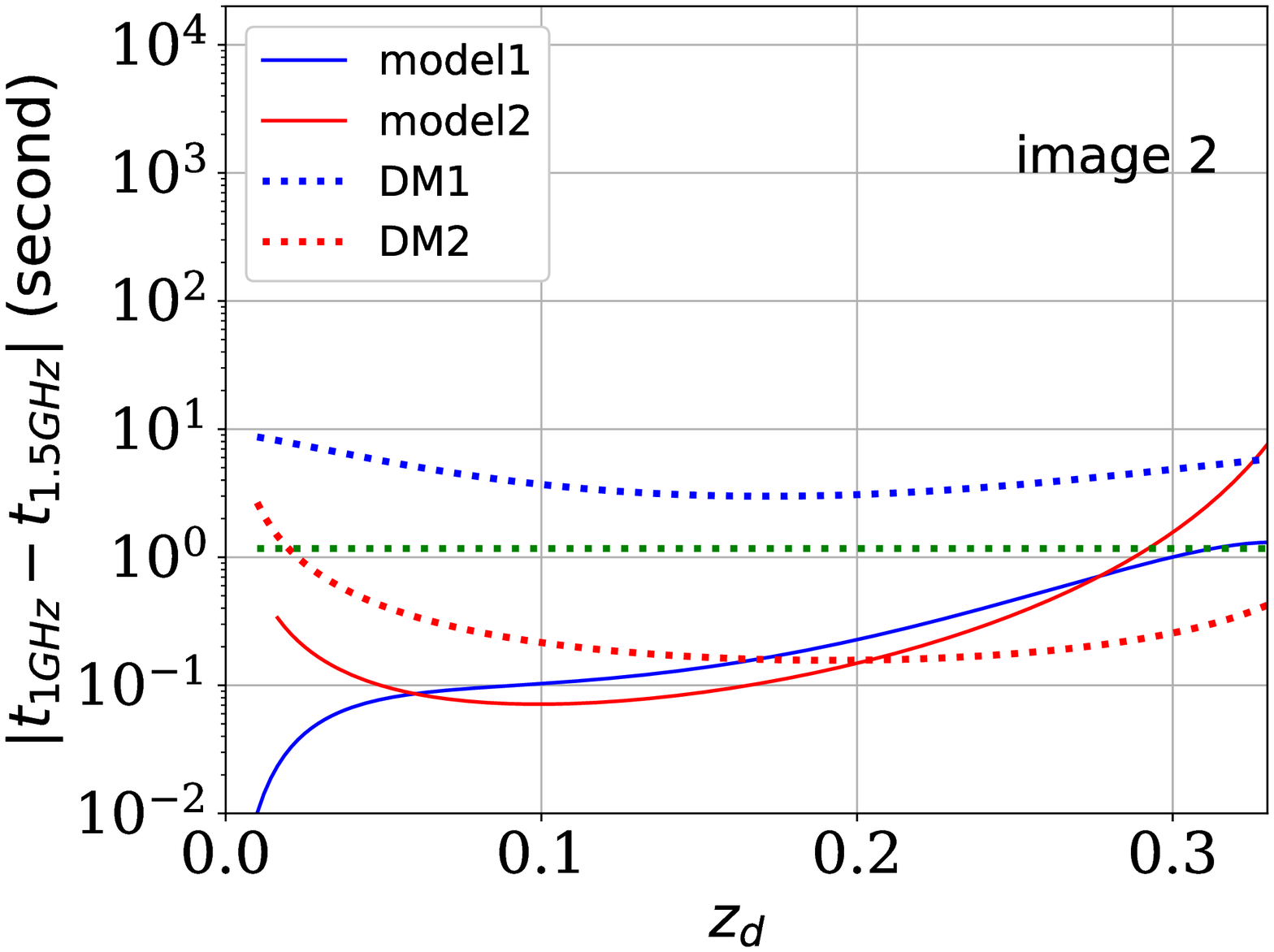}
\caption{The time delay as a function of the lens redshift $z_d$. The same source redshift ($z_s=0.5534$) is used. The solid curves are the gravitational lensing delay (first and second terms in Eq.\,\ref{eq:totaldelay}), and the dotted curves are the dispersion delay. The green dotted curves represent the dispersion delay caused by the plasma in IGM.}
\label{fig:td-zd}
\end{figure}

The plasma can exist all the way between the FRB and the observer, e.g. in the Milky Way, IGM or host galaxy, and all of them will cause the dispersion delay which only weakly depends on their redshifts. In many studies, it is necessary to distinguish the distances of the plasma, but this may be difficult in reality. On the other hand, the lensing geometric delay is strongly redshift dependent. 
We thus compare the dispersion delay with the lensing time delay as a function of the lens redshift (Fig.\,\ref{fig:td-zd}). We subtract the dispersion delay in the lensing time delay for better comparison. The same lens model and source redshift are adopted. A different source position ($\beta=(0.45,0.8)$ arcsec) is used in order to have two images at all lens redshifts. The two image positions in the two models move from $(2.876,2.697)$, $(-0.638,-1.939)$ to $(1.189,1.656)$, $(-0.079,-0.176)$ respectively. In both images, the delays generated by model 1 are small, and can be neglected in most cases. 
In model 2 the lensing delay is significant and cannot be neglected in most cases. In image 2 the lensing delays are greater than the dispersion delay at high lens redshifts. In image 1, the lensing delay also becomes important at high-z.
Besides the plasma in the lens galaxy, the IGM in the universe cannot be neglected, especially at high redshift. We thus adopt the empirical DM-z relation \citep{2020Natur.581..391M,2022MNRAS.509.4775J} to estimate the plasma along the line of sight to the source redshift. The green line in Fig.\,\ref{fig:td-zd} shows the delay difference caused by the plasma in IGM.
In some additional tests, the lensing delay can dominate the total time delay in a reasonable high plasma model.

\section{Summary and Discussions}
The Fast Radio Bursts (FRBs) are short duration radio transients with large discovery space. Strongly lensed FRBs are estimated to be significant in number in future surveys \citep{2022arXiv220614310C}, and will be detected in the near future. Such events can provide rich information in the study of cosmology and astrophysics due to high precision estimations of time delay. 
The plasma in the lens galaxy can cause extra deflections to the FRB signal. Unlike the gravitational lensing, plasma lensing shows frequency dependence. Such a frequency variation may not be detectable in image positions, but possible in the magnification and definitely in the time delay. The total time delay caused by the gravitational and plasma lensing can diverge significantly from the conventional plasma dispersion relation. In particular, when the deflection by either gravity or plasma is large, the geometric time delay can dominate the dispersion relation. The exact situation can be complicated, and depends on several factors, e.g. the mass distribution of the lens galaxy, the plasma density, and also the redshift. When the lens is close to us or the source, there is a high probability that the lensing delay plays a main role in the total frequency-delay relation.

The time delay relation provides a potential way to search for the lensed FRBs, e.g., one may search from exotic dynamic spectra of the FRBs \citep[e.g.][]{2021arXiv210713549T,2022RAA....22f5017W}. To do this, it is also necessary to take into account of the plasma lensing delay in the predictions of the applications of lensed FRBs. Moreover, the lensing FRBs will provide information in several studies, such as the plasma distribution in the lens galaxy. The macro-structure of plasma in the host galaxy of a lensed FRB can cause large refraction geometric time delay \citep[][]{2022MNRAS.509.5872E}, which will show up in all the multiple images. Such a feature caused by the plasma in IGM or Milky Way will not be present in all the lensed images, and may be used to study the distribution of plasma along the line of sight.

The gravitational lensing effect is achromatic, thus the frequency dependent time delay can only be caused by plasma. Moreover, the plasma lensing delay has a frequency dependence to the power of -4, which can be distinguished from the dispersion delay (-2). The mixture effect of gravity and plasma lensing can increase or decrease the frequency-dependent time delay. The intrinsic frequency-dependent time delay of the source can be degenerate with lensing effects, but the difference between multiple images can only be caused by lensing. The magnification changed by the macro-scale structure of plasma in the lens galaxy is small in most of our models. However, there is one possible exception. When the lensed image appears close to the centre of the lens, the image will be strongly de-magnified in gravitational lensing without plasma. Since the plasma deflection is opposite to that by gravity, this weakens the de-magnification. The un-observable central image (host galaxy image) may become observable (the lensed FRB image) in this case.

The exact properties of the lensed FRBs can be more complicated in several aspects. First, we only adopt a macro density model of the electrons in the lens galaxy. As we have shown, the density gradients (both mass and plasma) are important to determine the delay properties. The macro lensing model can easily change the total time delay, but difficult to account for the properties of all multiple images. However, sub-halos and small compact objects, such as stars and black holes will cause microlensing effects and can significantly change the lensing magnification and time delay \citep{2021ApJ...912..134C,2021ApJ...923..117C}.
Such microlensing effects may be degenerate with the macro lensing, but can be studied if there are multiple images. Plasma lensing from small scale structures will be difficult to detect \citep[e.g.][]{FRBplasma1,Tuntsov2016}, and is worth further investigations. On the other hand, the large scale environment, i.e. external shear and convergence, will also increase the time delay scale significantly, since the mixture delay from gravitational and plasma delay depends on the total lensing shear. Moreover, the gravity of the plasma (not only electrons) can also cause further deflections which is not included in current calculations. 
Thus, further detailed analysis is necessary for the searching, predictions and applications of the strongly lensed FRBs.

\section*{Acknowledgements}
We thank the referee for valuable comments which clarify several ambiguities. XE is supported by the NSFC Grant No.\,11873006 and the China Manned Space Project with No.\,CMS-CSST-2021-A12. SM is supported by the National Key Research and Development Program of China No. 2018YFA0404501, and NSFC Grant No. 11821303, 11761131004 and 11761141012. 
\section*{Data Availability}
The data underlying this article will be shared on reasonable request to the corresponding author.

\bibliographystyle{mnras}
\bibliography{lens,frb,plasmalens,stronglens,lensfrb}

\bsp	
\label{lastpage}
\end{document}